\newcommand{\PRE}[1]{{#1}} 
\documentclass[12pt,aps,prd,preprint,tightenlines,nosuperscriptaddress,
amsmath,amssymb,
nofootinbib]{revtex4}

\usepackage{enumerate}
\usepackage{color}
\usepackage[pdftex]{graphicx}
\usepackage[colorlinks=true,citecolor=blue,urlcolor=blue]{hyperref}
\graphicspath{{paperfigures/}}	        

\newcommand{\mev}{\text{MeV}}
\newcommand{\gev}{\text{GeV}}

\newcommand{\m}{\text{m}}

\renewcommand{\eqref}[1]{Eq.~(\ref{#1})}
\newcommand{\eqsref}[2]{Eqs.~(\ref{#1}) and (\ref{#2})}
\newcommand{\secref}[1]{Sec.~\ref{sec:#1}}

\newcommand{\figref}[1]{Fig.~\ref{fig:#1}}
\newcommand{\figsref}[2]{Figs.~\ref{fig:#1} and \ref{fig:#2}}

\newcommand{\slas}[1]{\! \not{\! \! #1}}
\newcommand{\slass}[1]{\! \not{\! #1}}

\begin{document}

\preprint{UCI-TR-2017-06}

\title{\PRE{\vspace*{1.5in}}
Impact of Resonance on Thermal Targets for  \\
Invisible Dark Photon Searches
\PRE{\vspace*{.4in}}}

\author{Jonathan L.~Feng\footnote{jlf@uci.edu}}
\affiliation{Department of Physics and Astronomy, University of
  California, Irvine, California 92697-4575 USA
\PRE{\vspace*{.4in}}}

\author{Jordan Smolinsky\footnote{jsmolins@uci.edu}}
\affiliation{Department of Physics and Astronomy, University of
  California, Irvine, California 92697-4575 USA
\PRE{\vspace*{.4in}}}


\begin{abstract}
\PRE{\vspace*{.2in}} Dark photons in the MeV to GeV mass range are important targets for experimental searches.  We consider the case where dark photons $A'$ decay invisibly to hidden dark matter $X$ through $A' \to XX$.  For generic masses, proposed accelerator searches are projected to probe the thermal target region of parameter space, where the $X$ particles annihilate through $XX \to A' \to \text{SM}$ in the early universe and freeze out with the correct relic density. However, if $m_{A'} \approx 2m_X$, dark matter annihilation is resonantly enhanced, shifting the thermal target region to weaker couplings.  For $\sim 10\%$ degeneracies, we find that the annihilation cross section is generically enhanced by four (two) orders of magnitude for scalar (pseudo-Dirac) dark matter.  For such moderate degeneracies, the thermal target region drops to weak couplings beyond the reach of all proposed accelerator experiments in the scalar case and becomes extremely challenging in the pseudo-Dirac case.  Proposed direct detection experiments can probe moderate degeneracies in the scalar case.  For greater degeneracies, the effect of the resonance can be even more significant, and both scalar and pseudo-Dirac cases are beyond the reach of all proposed accelerator and direct detection experiments.  For scalar dark matter, we find an absolute minimum that sets the ultimate experimental sensitivity required to probe the entire thermal target parameter space, but for pseudo-Dirac fermions, we find no such thermal target floor.  
\end{abstract}


\maketitle

\section{Introduction}
\label{sec:intro}

The universe appears to be filled with dark matter with a relic density of $\Omega_X h^2 = 0.1199 \pm 0.0022$, where $\Omega_X$ is the energy density of dark matter in units of the critical density, and $h \simeq 0.67$ is the reduced Hubble parameter~\cite{Ade:2015xua}.  Because the relic density is an important quantity and so precisely known, or perhaps because so little else is known about dark matter, scenarios in which dark matter is produced through a simple mechanism that gives the correct $\Omega_X$ attract special attention.   In particular, dark matter that begins in thermal equilibrium with the standard model and then freezes out with the correct thermal relic density is often considered especially well motivated.  Examples include weakly-interacting massive particles (WIMPs), weak-scale particles with weak interactions, and WIMPless dark matter~\cite{Feng:2008ya}, hidden sector particles that are lighter and more weakly-coupled than WIMPs (or heavier and more strongly-coupled than WIMPs), but nevertheless also have the correct thermal relic abundance.  For both WIMP and WIMPless dark matter, the region of parameter space that yields the correct thermal relic density, often known as the ``thermal target,'' provides a useful goal for current and proposed experimental searches.  

Dark photon models~\cite{Okun:1982xi,Holdom:1985ag,Boehm:2003hm,Pospelov:2008zw} are a simple and elegant realization of the WIMPless possibility.  Dark photons $A'$ are light gauge bosons that have coupling $g_X$ to dark matter $X$ in the hidden sector and couplings $\kappa q_f$ to standard model particles $f$, where $\kappa$ is the kinetic mixing parameter, and $q_f$ is the electric charge of $f$.  The dark matter's relic density is determined by the annihilation process $XX \to A' \to \text{SM}$, and for particular choices of $m_X$, $m_{A'}$, $g_X$, and $\kappa$, this annihilation process yields the correct thermal relic density. 

Typically, one considers $g_X \sim 1$ and $\kappa \ll 1$, where $\kappa$ is assumed to be suppressed, because it is generated at loop level.  The dark photon scenario then splits into two cases.  If $m_{A'} < 2 m_X$, dark photons always decay to the visible (standard model) sector.  In this case, dark photons are produced at accelerators through their interactions with standard model particles and decay back to standard model particles.  They mediate a new force, a revolutionary discovery in and of itself, but their implications for dark matter are not directly probed by accelerator experiments.  

If $m_{A'} > 2 m_X$, however, dark photons produced at accelerators typically decay invisibly to the hidden sector through $A' \to X X$.   In this case, experiments that produce dark photons also produce dark matter, and the signature is missing mass, energy, or momentum.  Of course, there is a long road ahead to identify the missing particle with the dark matter that permeates the universe, but at least in this case, the underlying process involves a dark matter candidate.  The number and variety of experiments that are potentially sensitive to invisibly-decaying dark photons is staggering.  They include BaBar~\cite{Lees:2014xha,Lees:2017lec}, CRESST II~\cite{Angloher:2015ewa}, E137~\cite{Batell:2014mga,Izaguirre:2017bqb}, LSND~\cite{Athanassopoulos:1996ds,deNiverville:2011it,deNiverville:2016rqh,Izaguirre:2017bqb}, and NA64~\cite{Banerjee:2016tad}, which currently bound various regions of parameter space, and BDX~\cite{Battaglieri:2016ggd,Bondi:2017gul,Izaguirre:2017bqb}, Belle II~\cite{HeartyBelle2}, COHERENT~\cite{CoherentWeb,deNiverville:2015mwa},  DarkLight~\cite{Balewski:2014pxa}, LDMX~\cite{LDMXWeb}, MiniBoone~\cite{deNiverville:2011it,deNiverville:2016rqh,Aguilar-Arevalo:2017mqx,Izaguirre:2017bqb}, MMAPS~\cite{MMAPS}, NA64~\cite{na64future},  PADME~\cite{Raggi:2014zpa,Raggi:2015gza}, SHiP~\cite{Alekhin:2015byh,Anelli:2015pba}, SBNe/SBN$\pi$~\cite{Antonello:2015lea,SBNRVW}, and VEPP-3~\cite{Wojtsekhowski:2012zq}, which will probe this scenario in the future.  The promise of discovering a portal to the dark sector in these experiments is significant, especially in the case of LDMX, which has been projected to probe all of the thermal target region for $\mev < m_X \alt \gev$, $m_{A'} \agt 3m_X$, and $g_X \sim 1$~\cite{Izaguirre:2015yja}.  In addition, a large number of direct detection experiments, which we discuss below, although not creating real dark photons, also probe these scenarios through $X \, \text{SM} \to X \, \text{SM}$ mediated by a $t$-channel $A'$, where the $A' X X$ interaction is the same one that mediates the dark photon's invisible decays.

The invisible decay case with $m_{A'} > 2 m_X$ is, however, also the ``half'' of parameter space in which dark matter annihilation may be resonantly enhanced.  This has been noted previously, for example, in Ref.~\cite{Izaguirre:2015yja} (see, in particular, the supplementary material), but the impact of the $A'$ resonance has not been investigated in detail.  For degeneracies $0 < m_{A'} - 2 m_X \alt T_f$, where $T_f$ is the temperature at freezeout, the kinetic energy of dark matter particles can put the annihilation process $XX \to A' \to \text{SM}$ on resonance.  As we will see, for $\sim 10\%$ degeneracies between $m_{A'}$ and $2 m_X$, this resonance may have an extraordinary effect.  For example, for scalar dark matter, the resonance generically raises the thermally-averaged annihilation cross section by four orders of magnitude.  To compensate this kinematic enhancement, the thermal relic density may be corrected by lowering $\kappa^2$ by four orders of magnitude, but the resulting thermal target region of parameter space is then beyond the reach of any proposed accelerator experiment.  Greater degeneracies move the thermal target to even weaker couplings.  

In the following sections, we consider dark photon models with both scalar and pseudo-Dirac dark matter, estimate the effect of the resonance analytically, and present numerical results for the impact of the resonance on the thermal target region for a wide range of mass degeneracies.

\section{Dark Photon Model}
\label{sec:darkphoton}

The dark photon model we consider is standard, but we present it here to establish notation and conventions.  The hidden sector has a broken U(1) symmetry, and its massive gauge boson kinetically mixes with the standard model photon.  In the mass basis, the resulting Lagrangian is
\begin{align}
	\mathcal L &=
	-\frac 14 F_{\mu\nu}F^{\mu\nu}
	-\frac 14 F'_{\mu\nu}F'^{\mu\nu}
	+ \frac{1}{2} m_{A'}^2 A'^2
	+ \sum_f \bar{f} (i \slass{\partial} - e q_f \slas{A} - \kappa e q_f \slas{A'} - m_f)  f \ ,
	\label{eq:Lagrangian}
\end{align}
where $F_{\mu\nu}$ and $F'_{\mu\nu}$ are the field strengths of the photon $A$ and dark photon $A'$, respectively, $m_{A'}$ is the dark photon's mass, $\kappa$ is the kinetic mixing parameter (we reserve $\epsilon$ for another quantity below), and $f$ are standard model fermions with electric charges $q_f$ and masses $m_f$.  

The dark photon may decay to $e^+e^-$ pairs throughout the parameter space we study.  The decay width is
\begin{equation}
\Gamma_e \equiv \Gamma (A' \to e^+ e^-) 
=  \frac{\kappa^2 e^2 m_{A'}}{12 \pi} 
\Biggl[ 1 - \left( \frac{2 m_e} {m_{A'}} \right)^{\! \! 2} \ \Biggr]^{1/2} 
\Biggl[ 1 + \frac{2 m_e^2}{m_{A'}^2} \Biggr] .
\end{equation}
For $m_{A'} > 2 m_{\mu}$, decays to muons and a number of hadronic states are also possible.  The full standard model decay width is 
\begin{align}
\Gamma_{\text{SM}} = \frac{\Gamma_e }{B_e(m_{A'}) } \ ,
\end{align}
where $B_e(m_{A'})$ is the branching fraction to $e^+e^-$ pairs of a dark photon with mass $m_{A'}$, which may be extracted from measurements of $\sigma (e^+ e^- \to e^+ e^-)/\sigma_{e^+ e^-}^\text{tot}$~\cite{Buschmann:2015awa}. 

The hidden sector also contains the dark matter.  We will consider both complex scalar~\cite{Boehm:2003hm} and pseudo-Dirac dark matter~\cite{TuckerSmith:2001hy,Izaguirre:2015zva,Izaguirre:2017bqb}.  For the scalar case, the dark matter Lagrangian is 
\begin{align}
	{\cal L}_\phi = \left| ( \partial_\mu + i g_X A'_{\mu} ) \phi \right|^2 - m_{\phi}^2 \left| \phi \right|^2  \ ,
\end{align}
where $\phi$ is the scalar dark matter particle with mass $m_{\phi}$, and $g_X = \sqrt{4 \pi \alpha_X}$ is the hidden sector gauge coupling.  In this case, the invisble (hidden sector) decay width is
\begin{align}
\Gamma_\phi \equiv \Gamma ( A' \to \phi \phi) 
= \frac{g_X^2 m_{A'}}{48 \pi} \Biggl[ 1 - \left(\frac{2 m_\phi}{m_{A'}} \right)^{\! \! 2} \ \Biggr]^{3/2} \ .
\end{align}

For the pseudo-Dirac case, we consider a dark photon that couples to two hidden Weyl fermions that have a Dirac mass and small, identical Majorana masses. In the mass basis, the resulting dark matter Lagrangian is
\begin{align}
	{\cal L}_\chi = \sum_{i = 1, 2} \bar{\chi}_i (i \slass{\partial} - m_i) \chi_i 
	- \left( g_X \bar{\chi}_2 \slas{A'} \chi_1 + \text{h.c.} \right) \ ,
\end{align}
where the two fermions $\chi_1$ and $\chi_2$ couple non-diagonally to the dark photon and have masses $m_1$ and $m_2$, respectively,  with a small mass splitting $\Delta \equiv m_2 - m_1$.   Below we will typically refer to $\chi_1$ as the dark matter particle $X$ with mass $m_X$ and to $\chi_2$ as the excited state with mass $m_X + \Delta$.  In this pseudo-Dirac case, the invisible, hidden sector decay width is
\begin{equation}
\Gamma_{\chi} \equiv \Gamma (A' \to \chi_1 \chi_2)
= \frac{g_X^2 m_{A'}}{12 \pi} 
\Biggl[ 1 \! - \! \left( \frac{2m_X \! + \! \Delta}{m_{A'}} \right)^{\! \! 2} \Biggr]^{1/2} 
\Biggl[ 1 \! + \! \frac{(2m_X \! + \! \Delta)^2}{2 m_{A'}^2} \Biggr]
\Biggl[ 1 \! - \! \frac{\Delta^2}{m_{A'}^2} \Biggr]^{3/2} .
\end{equation}

\section{Relic Densities Near Resonance: Analytic Estimate}
\label{sec:analytic}

The formalism for treating dark matter annihilation near a resonance was developed long ago~\cite{Gondolo:1990dk,Griest:1990kh}.  In this section, we follow the method of Ref.~\cite{Gondolo:1990dk} to derive a simple analytic estimate for the effect of a resonance on the thermal relic density for $\sim$ 10\% degeneracies.  In \secref{numerical}, we will refine the standard treatment to improve its validity off resonance.  We then use these results to derive more precise numerical results for cases with both more and less degeneracy, which we present in \secref{results}.

The thermal relic abundance of a dark matter particle $X$ is 
\begin{equation}
\Omega_X h^2 = 8.77 \times 10^{-11}~\gev^{-2} \left[\overline{g_\text{eff}^{1/2}} \int_{x_f}^{x_0} \frac{\langle \sigma v \rangle}{x^2} dx \right]^{-1} \ ,
\label{omega}
\end{equation}
where $\langle \sigma v\rangle$ is the thermally-averaged annihilation cross section, $x_0 = m_X/T_0 = 4.26 \times 10^{12} \ (m_X/\gev)$, $T_0$ is the temperature now, and $x_f = m_X/T_f$, where $T_f$ is the freezeout temperature.  The freezeout temperature is found by solving the equation
\begin{equation}
\frac{63 \times 5^{1/2} x_f^{-1/2} e^{-x_f} g}{32 \pi^3} \frac{g_\text{eff}^{1/2}(x_f)}{h_\text{eff}(x_f)} m_X  m_\text{Pl}  \langle \sigma v\rangle = 1 \ .
\label{xf}
\end{equation}
In these equations, $g_\text{eff}(x)$ and $h_\text{eff}(x)$ are the effective numbers of degrees of freedom for energy and entropy density, respectively, $\overline{g_\text{eff}^{1/2}}$ is the typical value of $g_\text{eff}(x)$ between $x_0$ and $x_f$, and $g$ is the number of $X$ spin degrees of freedom. 

To determine the thermal relic density, then, we must determine $\langle \sigma v\rangle$.  In this section, we consider the simple case where the dark matter annihilates through the dark photon resonance $X X \to A' \to \text{SM}$.  For this case, it is convenient to define
\begin{eqnarray}
s_0 &\equiv& 4 m_X^2 \label{s0} \\
\epsilon &\equiv& (s - s_0) / s_0 \\
\epsilon_R &\equiv& (m_{A'}^2 - s_0 ) / s_0 \\
\gamma_R &\equiv& m_{A'} \Gamma_{A'} / s_0 \label{gammaR} \ ,
\end{eqnarray}
where $\epsilon$, $\epsilon_R$, and $\gamma_R$ are dimensionless quantities that represent the kinetic energy of the collision, the kinetic energy required to be on resonance, and the width of the resonance, respectively. With a slight abuse of notation, in cases where it matters, for example, for the very small values of $\epsilon_R$ that we will consider below, these definitions should be considered to be in terms of physical quantities rather than Lagrangian parameters, so, for example, loop corrections have been included in the masses in Eqs.~(\ref{s0})--(\ref{gammaR}).

In general, $\langle \sigma v\rangle$ must be evaluated numerically, but the formalism simplifies greatly with three approximations. First, if the dark matter freezes out while non-relativistic, $x_f \equiv m_X/T_f \gg 1$, the thermally-averaged annihilation cross section is approximately~\cite{Gondolo:1990dk}
\begin{equation}
\langle \sigma v \rangle_{\text{NR}}
= \frac{2 x^{3/2}}{\pi^{1/2}} \int_0^\infty \sigma v_\text{lab} 
\epsilon^{1/2} e^{-x \epsilon} d \epsilon \ ,
\end{equation}
where $\sigma$ is the annihilation cross section, and 
\begin{equation}
v_\text{lab} = \frac{2 \epsilon^{1/2} (1 + \epsilon) ^{1/2} } { 1 + 2 \epsilon}
\label{vlab}
\end{equation}
is the relative velocity of the incoming particles in the rest frame of one of them.  We have verified that $x_f \sim 15$ and the non-relativistic approximation is valid to $\sim 10\%$ throughout the regions of parameter space we consider.

Second, if we are sufficiently near the $A'$ resonance, so $x_f \epsilon_R \alt 1$ or $m_{A'} - 2 m_X \alt m_X / x_f$, we may take $\sigma$ to have the Breit-Wigner form
\begin{equation}
\sigma_{\text{BW}} = \frac{4 \pi \omega}{p^2} B_i B_f 
\frac{m_{A'}^2 \Gamma_{A'}^2} {(s - m_{A'}^2)^2 + m_{A'}^2 \Gamma_{A'}^2}
= \frac{4 \pi \omega}{m_X^2 \epsilon} B_i B_f 
\frac{\gamma_R^2} {(\epsilon - \epsilon_R)^2 + \gamma_R^2} \ ,
\end{equation}
where $\omega = (2S_{A'}+1)/(2S_X+1)^2$, $S_{A'} = 1$ is the dark photon's spin, $S_X$ is the dark matter's spin, and $B_i = B(A' \to X X )$ and $B_f = 1- B_i = B(A' \to \text{SM})$ are the branching fractions to the hidden and visible sectors, respectively.  

Third,  if the dark photon's couplings are sufficiently weak, so $\gamma_R \ll 1$ or $\Gamma_{A'} \ll m_{A'}$, we may use the narrow width approximation and the Breit-Wigner cross section becomes a delta function.  In the numerical analysis described below in \secref{results}, we have verified that, even for large $\alpha_X \sim 0.5$, the narrow width approximation gives thermally-averaged cross sections that are in agreement with the full result at the $ 10\%$ level for $\epsilon_R \sim 0.1$, improving to $1 \%$ agreement for $\epsilon_R \lesssim 0.01$. 

Given these three simplifications, the thermally-averaged annihilation cross section near a resonance at freezeout is~\cite{Gondolo:1990dk}
\begin{eqnarray}
\langle \sigma v \rangle _{\text{res}}
&\approx& \frac{16 \pi^{3/2} \omega}{m_X^2} x_f^{3/2} 
\gamma_R B_i B_f\frac{  (1 + \epsilon_R)^{1/2}}{1 + 2 \epsilon_R } e^{- x_f \epsilon_R} \\
&\approx& \frac{16 \pi^{3/2} \omega}{m_X^2} x_f^{3/2} 
\frac{\Gamma_{A'}}{2 m_X} B_i B_f e^{- x_f \epsilon_R}  \ ,
\label{resonance}
\end{eqnarray}
where in the last step we have used the fact that $x_f \sim 20$, and so the ``near resonance'' assumption implies $\epsilon_R \ll 1$.  Since we are considering invisible decay scenarios, $B_i \approx 1$.  We may also write $\Gamma_{A'} B_f = \Gamma_f \equiv N_f \kappa^2 e^2 m_X / 12 \pi$, where $N_f$ is the effective number of kinematically accessible standard model decay channels.  We then find that
\begin{equation}
\langle \sigma v \rangle _{\text{res}} \approx
 \frac{2 \pi^{1/2} \omega}{3 m_X^2} x_f^{3/2} \kappa^2 e^2 N_f e^{- x_f \epsilon_R} \ .
\label{resonance3}
\end{equation}

In the absence of resonances, assuming $m_X \sim m_{A'}$, the typical value for the thermally-averaged annihilation cross section is
\begin{equation}
\langle \sigma v \rangle_{\text{non-res}} 
\sim \frac{\pi \kappa^2 \alpha \alpha_X}{m_X^2} \frac{1}{x_f^L}
= \frac{\kappa^2 e^2 g_X^2}{16 \pi m_X^2} \frac{1}{x_f^L} \ ,
\end{equation}
where $L=0 \ (1)$ for $s$-wave ($p$-wave) annihilation. The nearby resonance therefore enhances the thermally-averaged annihilation cross section by a factor
\begin{equation}
\frac{\langle \sigma v \rangle_{\text{res}}}
{\langle \sigma v \rangle_{\text{non-res}} }
\sim \frac{32 \pi^{3/2} \omega N_f }{3 g_X^2} x_f^{3/2} x_f^L e^{- x_f \epsilon_R}
\sim  \text{5,000} \ \frac{\omega N_f x_f^L e^{- x_f \epsilon_R}}{ g_X^2 } \ .
\end{equation}
We see that a resonance may enhance the annihilation cross section (and suppress the thermal relic density) by four (two) orders of magnitude for the case of $p$-wave ($s$-wave) annihilators when $m_{A'}$ and $2m_X$ are degenerate to $\sim 10\%$. 

This conclusion for the thermal relic density assumes $\Omega_X h^2 \sim \langle \sigma v \rangle ^{-1}$, as typically follows from \eqref{omega}.  This is valid for the $\sim 10\%$ degeneracies discussed here, but as we will see in \secref{results}, for even greater degeneracies $\epsilon_R \ll 0.1$, there are additional effects that enhance the resonance effect further.

\section{Relic Densities Near Resonance: Numerical Analysis}
\label{sec:numerical}

In this section, we present our method for numerically evaluating the thermal relic density near resonance.  Our numerical results assume dark matter is non-relativistic at freezeout, but unlike the analytic estimate of \secref{analytic}, do not assume the resonance is nearby and do not assume the narrow-width approximation. In addition, we present results for both the scalar case and the pseudo-Dirac case.  The method is a generalization of the treatment presented in Ref.~\cite{Gondolo:1990dk}. 

The contribution of an $s$-channel resonance to the dark matter annihilation cross section can always be written in the form
\begin{equation}
\sigma v_\text{lab} = F(\epsilon) \frac{m_{A'} \Gamma_{A'}}{(s-m_{A'}^2)^2 + m_{A'}^2 \Gamma_{A'}^2} \ ,
\label{eq: sigmavdecomp}
\end{equation}
where $v_{\text{lab}}$ is given in \eqref{vlab} and the dimensionless analytic function $F(\epsilon)$ encodes the cross section's dependence on the dimensionless kinetic energy $\epsilon \equiv (s-s_0)/s_0$, where $s_0 = 4 m_X^2$ in the scalar case and $s_0 = (2 m_X + \Delta)^2$ in the pseudo-Dirac case.

We can then exploit a special function to describe the terms of a partial cross section expansion in a compact and numerically well-described manner. In the non-relativistic thermal average
\begin{equation}
\langle \sigma v \rangle_\text{NR} = \frac{2 x^{3/2}}{\pi^{1/2}} \int_0^\infty \sigma v_\text{lab} \epsilon^{1/2} e^{-x \epsilon} d \epsilon \ ,
\end{equation}
we rewrite the integral as
\begin{equation}
\int_0^\infty \frac{1}{s_0} \text{Re}\left[\frac{i}{\epsilon_R + i \gamma_R - \epsilon} F(\epsilon) \epsilon^{1/2} e^{-x \epsilon}\right] d \epsilon \ .
\end{equation}
Substituting the Taylor expansion $F(\epsilon) = \sum_{\ell = 0}^{\infty} F^{(\ell)} \epsilon^\ell / \ell !$, we find
\begin{equation}
\langle \sigma v \rangle_\text{NR} = \frac{2 x^{3/2}\pi^{1/2}}{s_0} \sum_{\ell =0}^{\infty} \frac{ F^{(\ell)}}{\ell!} \text{Re}\left[\frac{i}{\pi}\int_0^\infty \frac{\epsilon^{\ell+1/2} e^{-x \epsilon}}{z_R - \epsilon} d\epsilon\right] \ ,
\end{equation}
where $z_R \equiv \epsilon_R + i \gamma_R$. The $s$- and $p$-wave terms of the above expansion can be written compactly as
\begin{eqnarray}
\langle \sigma v \rangle_\text{NR} &\approx& \frac{2 x^{3/2}\pi^{1/2}}{ s_0} \Bigl\lbrace F^{(0)} \text{Re}\left[z_R^{1/2} w(x^{1/2}z_R^{1/2}) \right]  \nonumber \\
&& \qquad \qquad \quad + F^{(1)} \left(\gamma_R \pi^{-1/2} x^{-1/2} + \text{Re}\left[z_R^{3/2} w(x^{1/2} z_R^{1/2})\right]\right) \Bigr\rbrace \ ,
\label{sigmaNR}
\end{eqnarray}
where 
\begin{equation} 
w(z) \equiv \frac{2iz}{\pi} \int_0^{\infty} \frac{e^{-t^2}}{z^2 - t^2} dt 
= e^{-z^2} \left(1+\frac{2i}{\sqrt{\pi}} \int_0^z e^{t^2} dt \right) 
\label{faddeeva}
\end{equation}
is the Faddeeva function.  The first form in \eqref{faddeeva} is useful to derive \eqref{sigmaNR}, and the second form can be used to evaluate the function numerically and efficiently.  In this work, numerical calculations were performed with the SciPy library~\cite{scipy} using the method of Ref.~\cite{Zaghloul:2012:A9C:2049673.2049679} to evaluate the Faddeeva function close to the real axis.

We now turn to the specific models we consider in this paper. For the case of scalar $X$ annihilating to standard model final states through $XX \to A' \to \text{SM}$, the cross section takes the form
\begin{equation}
\sigma v_\text{lab} = \frac{16 \pi \kappa^2  \alpha \alpha_X }{3 \left((s-m_{A'}^2)^2 + m_{A'}^2 \Gamma_{A'}^2\right)} \frac{\epsilon \left[m_e^2 + 2(1+\epsilon) m_X^2\right]\sqrt{1+\epsilon - m_e^2/m_X^2}}{(1+2\epsilon) \sqrt{1+\epsilon} ~B_e (2 m_X\sqrt{1+\epsilon})} \ ,
\end{equation}
which implies 
\begin{equation}
F(\epsilon) = \frac{16 \pi \kappa^2  \alpha \alpha_X }{3 m_{A'} \Gamma_{A'}} \frac{\epsilon \left[m_e^2 + 2(1+\epsilon) m_X^2\right]\sqrt{1+\epsilon - m_e^2/m_X^2}}{(1+2\epsilon) \sqrt{1+\epsilon} ~B_e (2 m_X\sqrt{1+\epsilon})} \ .
\end{equation}
This cross section is $p$-wave suppressed, as indicated by the leading factor of $\epsilon$. We therefore know that $F^{(0)}$ vanishes, and we can verify that
\begin{equation}
F^{(1)} = \frac{16 \pi \kappa^2  \alpha \alpha_X}{3 m_{A'} \Gamma_{A'}} \frac{ \left(m_e^2 + 2 m_X^2\right)\sqrt{1 - m_e^2/m_X^2}}{B_e (2 m_X)} \ .
\end{equation}
Given this, the thermally-averaged cross section is given by \eqref{sigmaNR}, and the thermal relic density and freezeout temperature can be determined by \eqsref{omega}{xf}.

For the pseudo-Dirac case, the cross section for annihilation through $\chi_1 \chi_2 \to A' \to \text{SM}$ is
\begin{align}
	\begin{split}
		\sigma v_\text{lab} &= \frac{4 \pi \kappa^2 \alpha \alpha_X }{3 s_0 \left[(s-m_{A'}^2)^2 + m_{A'}^2 \Gamma_{A'}^2\right]} \\
		& \quad \times\frac{(3+2\epsilon) \left[(1+\epsilon)s_0 + 2m_e^2\right]\left[s_0(1+\epsilon) - 4m_e^2\right]^{1/2} \left[s_0(1+\epsilon) - \Delta^2\right]^{1/2}}{(1+\epsilon) (1+2\epsilon) ~B_e(\sqrt{s_0(1+\epsilon)})} \ ,
	\end{split}
\end{align}
which implies 
\begin{align}
	\begin{split}
		F(\epsilon) &= \frac{4 \pi \kappa^2 \alpha \alpha_X }{3 s_0 m_{A'} \Gamma_{A'}} \frac{(3+2\epsilon) \left[(1+\epsilon)s_0 + 2m_e^2\right]\left[s_0(1+\epsilon) - 4m_e^2\right]^{1/2} \left[s_0(1+\epsilon) - \Delta^2\right]^{1/2}}{(1+\epsilon) (1+2\epsilon) ~B_e(\sqrt{s_0(1+\epsilon)})} \ .
	\end{split}
\end{align}
This is an $s$-wave cross section, and it is easy to read off the constant term in the Taylor expansion of $F$:
\begin{equation}
F^{(0)} = \frac{4 \pi \kappa^2 \alpha \alpha_X }{s_0 m_{A'} \Gamma_{A'} B_e(\sqrt{s_0})} (s_0 + 2 m_e^2)(s_0 - 4m_e^2)^{1/2} (s_0 - \Delta^2)^{1/2} \ .
\end{equation}
This determines the leading contribution to the thermally-averaged cross section through \eqref{sigmaNR}.

There is an additional complication in the pseudo-Dirac case: when the dark sector consists of multiple nearly-degenerate species, we must include coannihilation factors in the thermally-averaged cross section~\cite{Griest:1990kh} and the freezeout condition.  The effective thermally-averaged cross section is
\begin{equation}
\langle \sigma v \rangle_\text{eff} = \frac{2 (1+\Delta/m_X)^{3/2} e^{-x \Delta/m_X}}{\left[1+(1+\Delta/m_X)^{3/2}e^{-x \Delta/m_X}\right]^2} \, \langle \sigma v \rangle _{\text{NR}} \ .
\end{equation}
The freezeout condition is modified to
\begin{equation}
\frac{63 \times 5^{1/2} x_f^{-1/2} e^{-x_f} g}{32 \pi^3} \frac{g_\text{eff}^{1/2}}{h_\text{eff}} m_X  m_\text{Pl} \left[1+ \left(1+ \frac{\Delta}{m_X}\right)^{3/2}e^{-x_f \Delta/m_X}\right] \langle \sigma v\rangle_\text{eff} = 1 \ ,
\end{equation}
and the relic abundance is given by
\begin{equation}
\Omega_X h^2 = 8.77 \times 10^{-11}~\gev^{-2} \left[\overline{g_\text{eff}^{1/2}} \int_{x_f}^{x_0} \frac{\langle \sigma v \rangle_\text{eff}}{x^2} dx \right]^{-1} \ .
\end{equation}

\section{Results}
\label{sec:results}

\begin{figure}[tbp] 
	\hspace*{-.5cm}
	\includegraphics[width=0.46\linewidth]{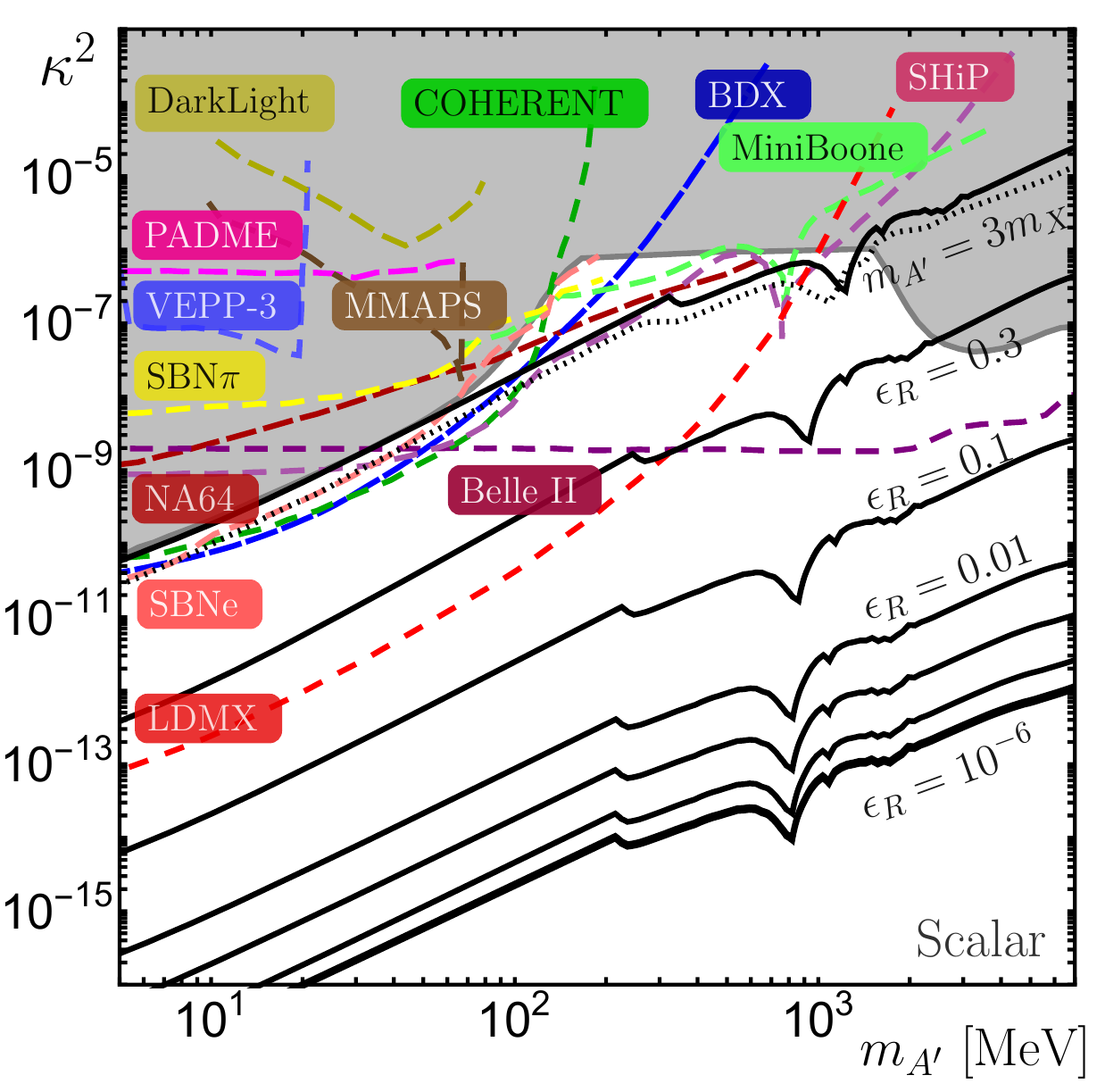} \\ 
	\includegraphics[width=0.46\linewidth]{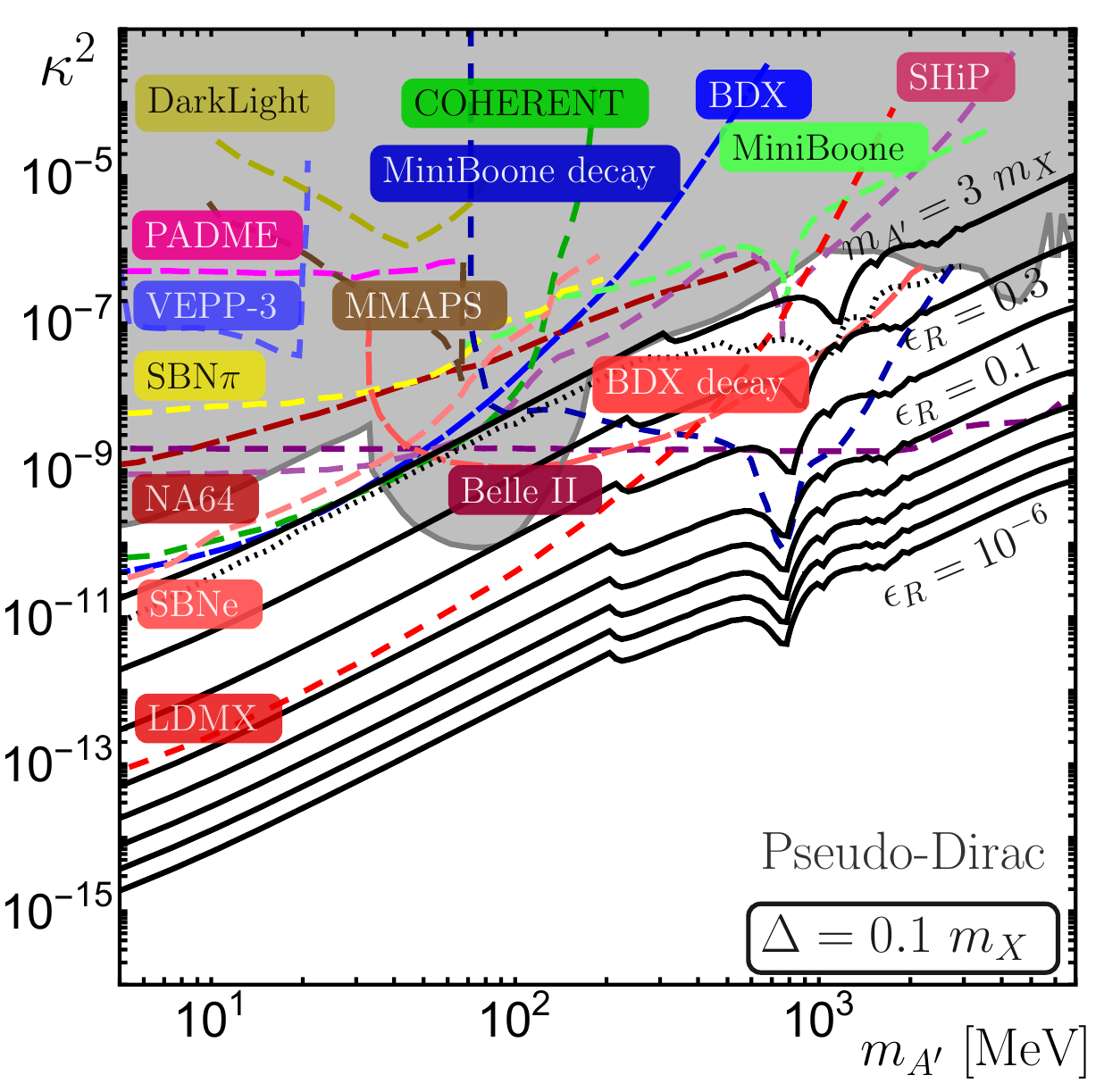} \quad
	\includegraphics[width=0.46\linewidth]{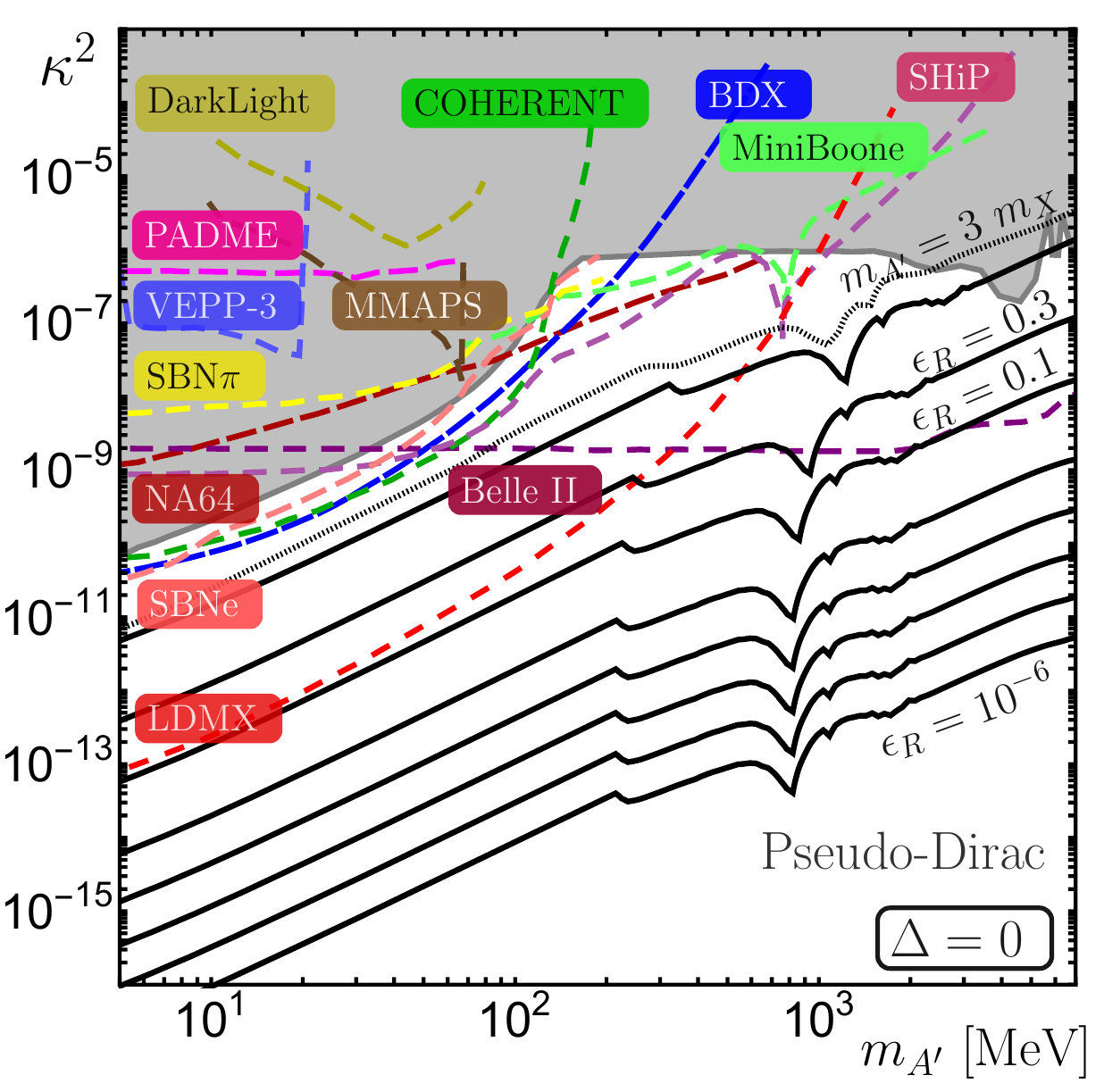} 
	\vspace*{-0.1in} 
	\caption{\textbf{Thermal targets and accelerator search experiments.}  \textsc{Solid Black Contours:} Thermal target contours in dark photon parameter space $(m_{A'}, \kappa^2)$ for $\alpha_X = 0.5$ in the non-degenerate cases with $m_{A'} = 3 m_X$ and $\epsilon_R \equiv (m_{A'}^2 - 4 m_X^2)/(4m_X^2)= 0.3$, and in the degenerate cases with $\epsilon_R = 10^{-n}$, where $n = 1, 2, \ldots 6$.  In the scalar case (\textsc{top}) the thermal relic contours reach a floor near $\epsilon_R = 10^{-6}$, where the thermal relic abundance requirement becomes inconsistent with the requirement that the dark photon decay is dominantly invisible. The $\epsilon_R = 10^{-5}$ contour is displayed in this plot, but is close enough to the $\epsilon_R = 10^{-6}$ contour that they appear to overlap. In the pseudo-Dirac case (\textsc{bottom}) the thermal relic contours extend to arbitrarily low values of $\kappa^2$ with appropriate choice of $\epsilon_R$, but we display only a small number of contours to avoid clutter. In the \textsc{bottom left} panel, the mass splitting is $\Delta = 0.1 \, m_X$, and these models evade direct detection bounds. In the \textsc{bottom right} panel, $\Delta = 0$, which illustrates the effect of decreasing $\Delta$ on the thermal target regions. \textsc{Dotted Black Contours}: Thermal target contours for $m_{A'}=3 m_X$ from relativistic treatments of freeze-out for the scalar~\cite{Izaguirre:2015yja} and pseudo-Dirac~\cite{Izaguirre:2017bqb} cases. \textsc{Gray Shaded:} Regions excluded by current bounds (see text). \textsc{Dashed Contours:} Projected reaches of proposed dark photon and dark matter accelerator searches (see text). }
	\label{fig:results}
	\vspace*{-0.1in}
\end{figure}

We present here the results of our analysis using the formalism of \secref{numerical}. Thermal relic contours for the benchmark, near-maximal perturbative value of the dark fine structure constant $\alpha_X = 0.5$ are shown in \figref{results} for various values of $\epsilon_R$. The contours for the non-degenerate case $\m_{A'} = 3 m_X$ agree within $\sim 10\%$ with those presented previously for scalar~\cite{Izaguirre:2015yja} and pseudo-Dirac~\cite{Izaguirre:2017bqb} dark matter, which are shown as dotted curves. The minor discrepancy is due to a $\sim 10\%$ difference between the thermally-averaged cross section in the non-relativistic approximation used here and the relativistic thermally-averaged cross section used in Refs.~\cite{Izaguirre:2015yja,Izaguirre:2017bqb} at freeze-out temperatures near $x_f = 15$.  For the degenerate case $\epsilon_R = 0.1$, as expected given the analytic estimate of \secref{analytic}, the thermal targets for scalar (pseudo-Dirac) dark matter move to values of $\kappa^2$ that are four (two) orders of magnitude lower, relative to the $\m_{A'} = 3 m_X$ non-degenerate case.  

The gray shaded regions in \figref{results} are excluded by various combinations of current constraints from BaBar~\cite{Lees:2014xha,Lees:2017lec}, CRESST II~\cite{Angloher:2015ewa}, E137~\cite{Batell:2014mga,Izaguirre:2017bqb}, LSND~\cite{Athanassopoulos:1996ds,deNiverville:2011it,deNiverville:2016rqh,Izaguirre:2017bqb}, and NA64~\cite{Banerjee:2016tad}.  The CRESST II bounds are not applicable to pseudo-Dirac dark matter with $\Delta = 0.1 \, m_X$, while exclusions from non-observation of excited state decays $\chi_2 \to \chi_1 e^+ e^-$ at E137 and LSND~\cite{Izaguirre:2017bqb} apply only to pseudo-Dirac dark matter with $\Delta > 2 m_e$.  Also included in \figref{results} are the projected sensitivities of planned accelerator-based dark photon and dark matter searches at BDX~\cite{Battaglieri:2016ggd,Bondi:2017gul,Izaguirre:2017bqb}, Belle II~\cite{HeartyBelle2}, COHERENT~\cite{CoherentWeb,deNiverville:2015mwa}, DarkLight~\cite{Balewski:2014pxa}, LDMX~\cite{LDMXWeb}, MiniBoone~\cite{deNiverville:2011it,deNiverville:2016rqh,Aguilar-Arevalo:2017mqx,Izaguirre:2017bqb}, MMAPS~\cite{MMAPS}, NA64~\cite{na64future}, PADME~\cite{Raggi:2014zpa,Raggi:2015gza}, SHiP~\cite{Alekhin:2015byh,Anelli:2015pba}, SBNe/SBN$\pi$~\cite{Antonello:2015lea,SBNRVW}, and VEPP-3~\cite{Wojtsekhowski:2012zq}.  The excluded regions and future sensitivities assume $m_{A'} = 3 m_X$.  For comparison with the degenerate case contours with $m_{A'} \approx 2 m_X$, these contours may be shifted up or down by ${\cal O}(1)$ factors.  For a comprehensive overview of existing constraints and future experimental sensitivities, see Ref.~\cite{Battaglieri:2017aum}.

Direct detection experiments can also probe these invisible dark photon scenarios.  Although not searches for dark photons per se, they probe the $A' X X$ vertex that induces invisible $A'$ decay through its role in inducing $X \, \text{SM} \to X \, \text{SM}$ scattering through a $t$-channel $A'$.  To facilitate comparison with direct detection experiments, it is convenient~\cite{Essig:2011nj,Essig:2015cda} to express the thermal relic parameter values in terms of
\begin{equation}
\bar{\sigma}_e = \frac{16 \pi \mu_{X,e}^2 \alpha \kappa^2 \alpha_X}{(m_{A'}^2 + \alpha^2 m_e^2)^2} \ ,
\end{equation}
where $\mu_{X,T}$ denotes the reduced mass of the dark matter-target system with $T = e$, nucleon, or nucleus. For the case of Majorana dark matter, the definition of $\bar{\sigma}_e$ includes an additional factor of $2 (\mu_{X,T}^2/m_X^2) v_X^2$, where $v_X = 10^{-3}$ is the characteristic DM halo velocity. 

\begin{figure}[tbp] 
	\hspace*{-.5cm}
	\includegraphics[width=0.46\linewidth]{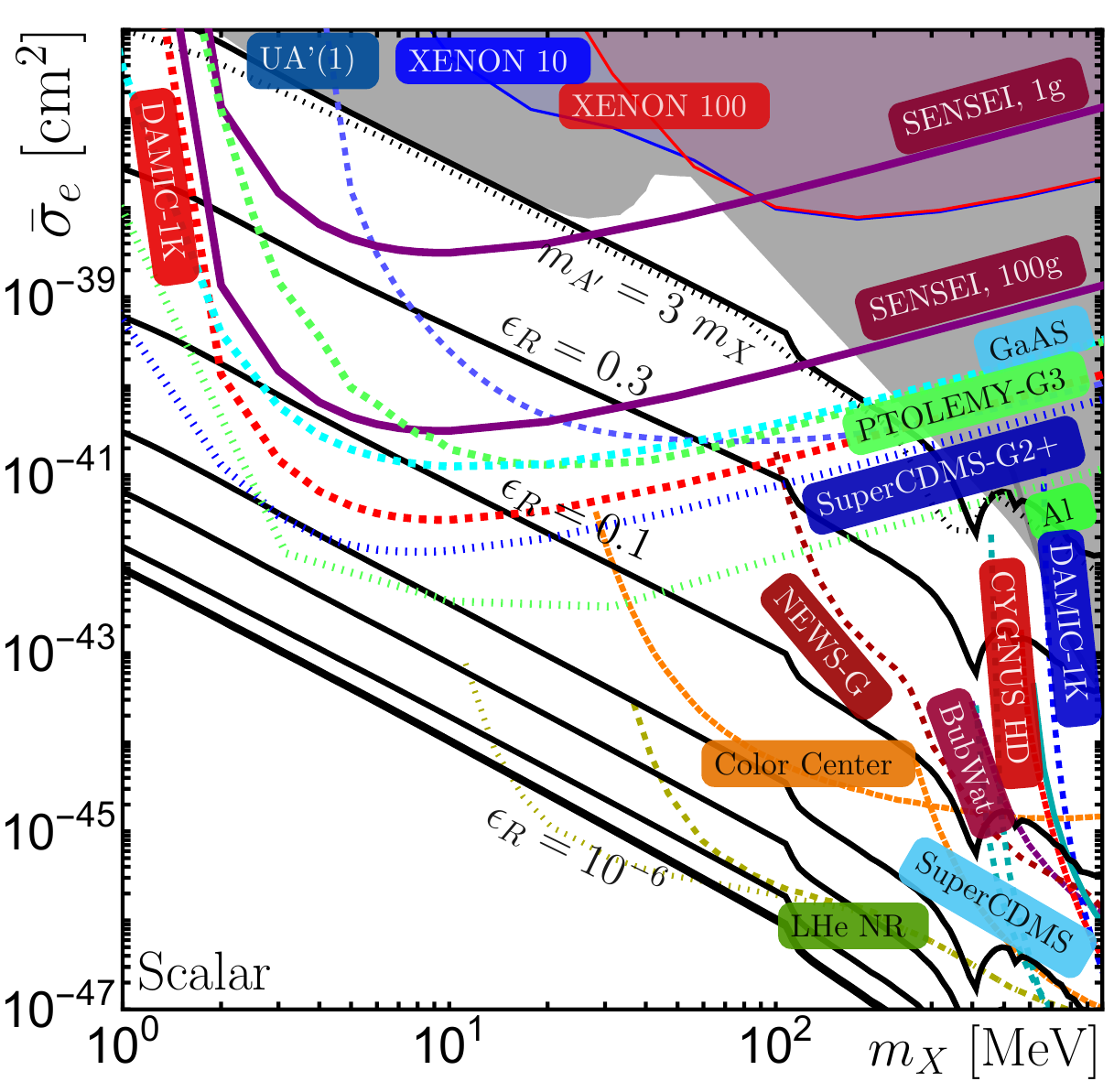} \quad
	\includegraphics[width=0.46\linewidth]{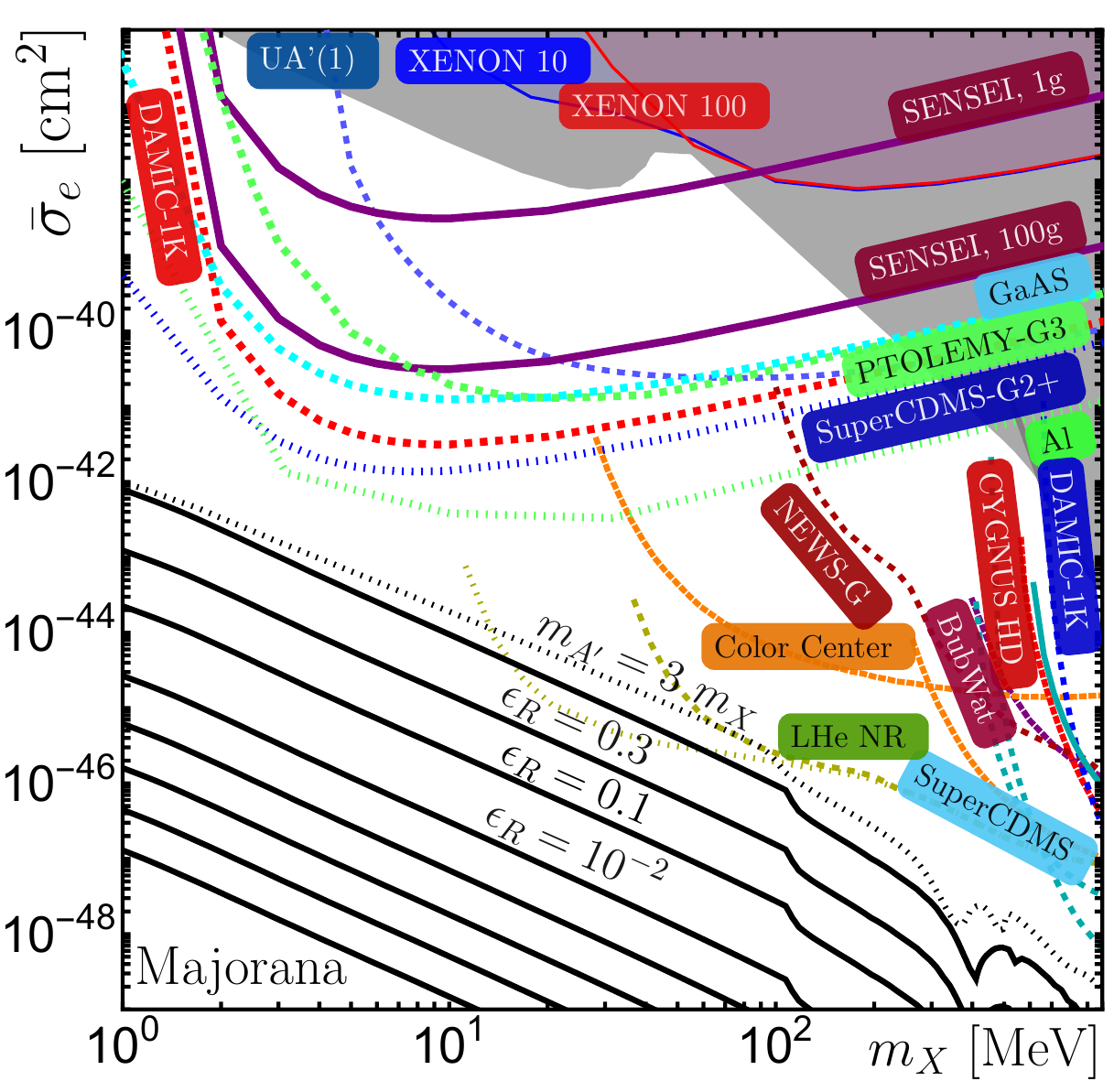}
	\vspace*{-0.1in} 
	\caption{\textbf{Thermal targets and direct detection search experiments.}  \textsc{Solid Black Contours:} Thermal target contours as in \figref{results}, but in the direct detection parameter space $(m_X, \bar{\sigma}_e)$, where $\bar{\sigma}_e = (16 \pi \mu_{X,e}^2 \alpha \kappa^2 \alpha_X)/(m_{A'}^2 + \alpha^2 m_e^2)^2$ (see text). In the scalar case (\textsc{left}) a resonantly annihilating thermal relic may still be within reach of future nuclear recoil experiments. In the \textsc{right} panel we show only the lines for Majorana DM-nuclear scattering, with $m_T = 0.936 ~\gev$; the thermal targets for DM-electron scattering scale approximately with the target mass squared, placing them far out of reach of current and proposed experiments.  \textsc{Dotted Black Contours}: Relativistic thermal relic contours, as in \figref{results}.  \textsc{Gray Shaded:} Regions excluded by current bounds (see text). \textsc{Colored Contours:} Projected reaches of proposed direct detection experiments (see text).}
	\label{fig:resultsDD}
	\vspace*{-0.1in}
\end{figure}

In \figref{resultsDD} we show the same thermal targets as in \figref{results}, but expressed in the $(m_X, \bar{\sigma}_e)$ parameter space and compared to current and proposed direct detection experiments. \figref{resultsDD} includes current exclusions from XENON \cite{Essig:2012yx,Essig:2017kqs}, as well as projected regions of sensitivity~\cite{Essig:2015cda,Essig:2012yx,Derenzo:2016fse,Hochberg:2016ntt,Hochberg:2015fth,Hochberg:2015pha} for CYGNUS HD-10, DAMIC-1K~\cite{Aguilar-Arevalo:2016ndq,Aguilar-Arevalo:2016zop}, NEWS, PTOLEMY-G3, SENSEI~\cite{Tiffenberg:2017aac}, SuperCDMS~\cite{Agnese:2014aze,Agnese:2015nto}, UA$'$(1), and future experiments based on GaAs scintillators~\cite{Derenzo:2016fse} superconducting aluminum, superfluid helium~\cite{Schutz:2016tid,Knapen:2016cue,Guo:2013dt,Car:2017,Maris:2017xvi}, color center production~\cite{Essig:2016crl,Budnik:2017sbu}, magnetic bubble chambers~\cite{Bunting:2017net}, scintillating bubble chambers~\cite{Baxter:2017ozv}, and bremsstrahlung in inelastic DM-nucleus scattering \cite{Kouvaris:2016afs,McCabe:2017rln}.  Exclusions and regions of sensitivity for nuclear recoil experiments are converted into limits and projected sensitivities for $\bar{\sigma}_e$ using
\begin{equation}
\bar{\sigma}_e = 4 \frac{\mu_{X, e}^2}{\mu_{X, N}^2} \sigma_N \ ,
\end{equation}
which makes it possible to compare the thermal targets and sensitivities of both electron and nuclear recoil experiments in the same parameter space. As in \figref{results}, the excluded regions and future sensitivities assume $m_{A'} = 3 m_X$ and are reviewed in Ref.~\cite{Battaglieri:2017aum}.

Comparing the thermal targets with the existing constraints and projected sensitivities, we see that for the scalar case and $\epsilon_R \sim 10\%$, the thermal target cannot be probed in any proposed accelerator or beam dump experiment.  For the fermionic dark matter cases and $\epsilon_R \sim 10\%$, the thermal target is also beyond the reach of all proposed accelerator-based experiments, with the exception of LDMX, for which it is at the border of sensitivity, and becomes very challenging for smaller mass splittings $\Delta$.  For direct detection experiments, the thermal target for moderate degeneracy $\epsilon_R \sim 10\%$ is still within the projected reach of some far future experiments in the scalar case, but is beyond all proposed experiments in the Majorana case.  

We also show results for smaller values of $\epsilon_R$ in \figref{results}.  For greater degeneracies, the thermal target region moves to even lower values of $\kappa^2$.  For $\epsilon_R \sim 10^{-6}$, for example, the thermal targets are essentially beyond all proposed accelerator and direct detection experiments for both the scalar and fermionic dark matter cases.   

At first sight, the extreme suppression of the preferred values of $\kappa^2$ might be surprising, since the thermally-averaged cross section, for example, in \eqref{resonance3}, becomes independent of $\epsilon_R$ for $x_f \epsilon_R \ll 1$.  However, for very small $\epsilon_R$, the dark matter continues to annihilate long after freezeout as the universe cools.  This is accounted for by the integral in \eqref{omega}, and that integral is sensitive to $\epsilon_R$, even it is very small.

To understand this behavior, it is convenient to use the narrow width approximation. In the case that $\Gamma_{A'} \ll m_{A'}$ we can write the generic resonant cross section as
\begin{equation}
\sigma v_\text{lab} \approx \frac{\pi}{s_0} F(\epsilon) \delta(\epsilon - \epsilon_R) \ ,
\end{equation}
which yields for the thermal average
\begin{equation}
\langle \sigma v \rangle_\text{NR} \approx \frac{2 \pi^{1/2} x^{3/2}}{s_0} \epsilon_R^{1/2} F(\epsilon_R) e^{-x \epsilon_R} \ .
\end{equation}
We notice that in both the scalar and pseudo-Dirac cases, as long as $\Gamma_{\text{SM}} \ll \Gamma_{\phi}, \Gamma_{\chi}$ and $\epsilon_R \ll 1$, the quantity $F(\epsilon_R)$ scales like $\epsilon_R^{-1/2}$. Therefore the thermally-averaged cross section's dependence on $\epsilon_R$ is contained entirely in the factor $\exp(-x \epsilon_R)$. This observation implies a simple relation between values of $\epsilon_R$ and $\kappa$ that yield the correct relic abundance:
\begin{equation}
\Omega_X h^2 \propto \left[\int_{x_f}^{x_0} \frac{\kappa^2}{x^{1/2}}e^{-x \epsilon_R} ~dx\right]^{-1} \propto \frac{\sqrt{\epsilon_R}}{\kappa^2} \ .
\end{equation}
We see that as $\epsilon_R \rightarrow 0$, a decrease of $\epsilon_R$ by an order of magnitude requires $\kappa^2$ to decrease by a factor of $\sqrt{10}$ to maintain the correct relic abundance. 

What are the smallest possible values of $\epsilon_R$?  In the scalar dark matter case, as we lower $\epsilon_R$, eventually the phase-space suppression of hidden sector decays will outweigh the kinetic mixing suppression of standard model decays, so that our assumption of invisible decays fails. Neglecting the electron mass, the requirement that the invisible width dominates implies
\begin{equation}
\Gamma_{\phi} \approx \frac{\alpha_X m_{A'}}{12} \epsilon_R^{3/2} \gtrsim \frac{ \alpha m_{A'}}{3 B_e}\kappa^2 \approx \Gamma_{\text{SM}} \ .
\end{equation}
It is clear, based on the power law dependence on $\epsilon_R$, that this condition cannot hold simultaneously with the thermal relic constraint over all of parameter space. For a given $\alpha_X$, there will be a minimum value of $\epsilon_R$ below which visible decays dominate the dark photon width. We find this minimum value to be $\epsilon_R^\text{min} \approx 10^{-6}$ for $\alpha_X = 0.5$.

In contrast, in the pseudo-Dirac case, the invisible width condition is
\begin{equation}
\Gamma_\chi \approx \frac{\alpha_X m_{A'}}{3} \epsilon_R^{1/2} \gtrsim  \frac{ \alpha m_{A'}}{3 B_e}\kappa^2 \approx \Gamma_{\text{SM}} \ ,
\end{equation}
which follows the same scaling as the thermal relic condition. In the pseudo-Dirac case, then, it is possible to lower the thermal target region to arbitrarily low values of $\kappa^2$ by choosing the dark matter to be arbitrarily close to resonance.  Put another way, enforcing the thermal relic constraint on a pseudo-Dirac dark sector ``accidentally'' fixes the ratio between the visible and invisible widths of the dark photon, so that the dual assumptions of mostly invisible dark photon decays and of thermal relic pseudo-Dirac dark matter may hold concurrently for all values of $\epsilon_R$. 

This is a remarkable result. It may be possible, in principle, to construct an experiment that can truly probe all of the thermal relic parameter space for perturbative theories of complex scalar dark matter coupled to a dark photon, but theories of fermionic dark matter may evade any such search by a fine-tuned choice of the dark sector masses. 

The interesting behavior for highly degenerate cases may have interesting consequences in other contexts.  For example, in the case of Kaluza-Klein dark matter~\cite{Servant:2002aq,Cheng:2002ej}, level-1 fermionic dark matter with mass $m_{\text{KK}}$ may annihilate through level-2 resonances with masses near $2 m_{\text{KK}}$, providing a rationale for high degeneracies.  These resonances will impact thermal relic density calculations~\cite{Kakizaki:2005en}, but for extreme degeneracies, our results imply that there may also be interesting astrophysical signals from dark matter annihilation long after freeze out.  Other interesting implications of resonances for such TeV-scale dark matter have been explored in Refs.~\cite{Feldman:2008xs,Ibe:2008ye,Guo:2009aj}.

\section{Conclusions}
\label{sec:conclusions}

The absence of the discovery of WIMPs and other classic dark matter candidates has motivated many new dark matter candidates in recent years.  Among those that are often seen as especially motivated are those that are in thermal equilibrium with the standard model at early times, but then freeze out with the correct thermal relic density.  The regions of model parameter space that give the desired relic density are ``thermal targets'' that provide important goals for new experimental searches.  

In this study, we considered dark photon scenarios in which the dark photon decays invisibly to dark matter through $A' \to XX$.  Such scenarios can be probed by experiments searching for missing mass, energy, or momentum.  For generic $A'$ and $X$ masses, proposed experiments, notably LDMX, are projected to be sensitive to the thermal target parameter space.  Direct detection experiments may also be sensitive to these scenarios by searches for $X \, \text{SM} \to X \, \text{SM}$ induced by $t$-channel $A'$ exchange.

Of course, in such scenarios, since $m_{A'} > 2 m_X$, the annihilation process $XX \to A' \to \text{SM}$ can be enhanced by the $A'$ resonance when the initial state dark matter particles have sufficient kinetic energies. In this work, we have found that for $m_{A'} - 2 m_X \sim 0.1 \, m_X$, the resonance implies a kinematic enhancement of the annihilation rate for scalar (pseudo-Dirac) dark matter of four (two) orders of magnitude, or, alternatively, the thermal target parameter space moves to values of the kinetic mixing $\kappa^2$ that are four (two) orders of magnitude smaller.   We derived these results using a simple analytic estimate in \secref{analytic} and through a more accurate numerical analysis in \secref{numerical}.  

The resulting thermal targets are shown in \figsref{results}{resultsDD}.  Even for the case of a 10\% degeneracy $m_{A'} - 2 m_X \sim 0.1 \, m_X$, we find that the thermal targets are very difficult to probe. For the scalar case, the thermal target is below the projected reach of LDMX and all other proposed accelerator experiments. For the pseudo-Dirac case, the thermal target is also beyond the reach of all proposed accelerator-based experiments, with the exception of LDMX, for which it is at the border of sensitivity, and becomes very challenging for smaller mass splittings.  Direct detection experiments do slightly better, as the thermal target for $\epsilon_R \sim 10\%$ is still within the projected reach of some far future experiments in the scalar case, but the thermal targets are still beyond all proposed experiments in the Majorana case.  

For even greater degeneracies, with $m_{A'} - 2 m_X \ll 0.1 \, m_X$, the thermal targets move to even lower values of $\kappa^2$.  For $\epsilon_R \sim 10^{-6}$, the thermal targets are essentially beyond all proposed accelerator and direct detection experiments for both the scalar and fermionic dark matter cases.   

Interestingly, for the case of scalar dark matter, for extreme degeneracies, the condition that the $A'$ decays dominantly invisibly becomes inconsistent with the thermal relic condition.  This establishes a floor at  $m_{A'} - 2 m_X \sim 10^{-6} \, m_X$ that is roughly four orders of magnitude in $\kappa^2$ below the projected reach of LDMX, but which is the ultimate goal for an experiment that can probe the entire thermal target region. The floor of the scalar dark matter parameter space may be accessible to future superfluid helium experiments. Unfortunately, for pseudo-Dirac dark matter, there is no such floor for the thermal target.  Of course, barring some more fundamental rationale, the fine-tuning required for such degeneracies is extreme, and ultimately other constraints will apply.

\acknowledgments

We are grateful to Anthony DiFranzo, Bertrand Echenard, Eder Izaguirre, Yoni Kahn,  Gopolang Mohlabeng, Brian Shuve, and, especially, Rouven Essig and Gordan Krnjaic for helpful discussions.  This work is supported in part by NSF Grant No.~PHY-1620638.   J.L.F. is supported in part by Simons Investigator Award \#376204, and J.S. is supported in part by Department of Education GAANN Grant No.~P200A150121 at UC Irvine. 

\bibliography{bibresonanceaprime}

\providecommand{\href}[2]{#2}\begingroup\raggedright\begin{thebibliography}{10}

\bibitem{Ade:2015xua}
{\bfseries Planck} Collaboration, P.~A.~R. Ade {\em et al.}, ``{Planck 2015
  results. XIII. Cosmological parameters},''
  \href{http://dx.doi.org/10.1051/0004-6361/201525830}{{\em Astron. Astrophys.}
  {\bfseries 594} (2016) A13},
\href{http://arxiv.org/abs/1502.01589}{{\ttfamily arXiv:1502.01589
  [astro-ph.CO]}}.

\bibitem{Feng:2008ya}
J.~L. Feng and J.~Kumar, ``{The WIMPless Miracle: Dark-Matter Particles without
  Weak-Scale Masses or Weak Interactions},''
  \href{http://dx.doi.org/10.1103/PhysRevLett.101.231301}{{\em Phys. Rev.
  Lett.} {\bfseries 101} (2008) 231301},
\href{http://arxiv.org/abs/0803.4196}{{\ttfamily arXiv:0803.4196 [hep-ph]}}.

\bibitem{Okun:1982xi}
L.~B. Okun, ``{Limits of electrodynamics: paraphotons?},'' {\em Sov. Phys.
  JETP} {\bfseries 56} (1982) 502.
[Zh. Eksp. Teor. Fiz. 83, 892 (1982)].

\bibitem{Holdom:1985ag}
B.~Holdom, ``{Two U(1)'s and Epsilon Charge Shifts},''
\href{http://dx.doi.org/10.1016/0370-2693(86)91377-8}{{\em Phys. Lett.}
  {\bfseries B166} (1986) 196--198}.

\bibitem{Boehm:2003hm}
C.~Boehm and P.~Fayet, ``{Scalar dark matter candidates},''
  \href{http://dx.doi.org/10.1016/j.nuclphysb.2004.01.015}{{\em Nucl. Phys.}
  {\bfseries B683} (2004) 219--263},
\href{http://arxiv.org/abs/hep-ph/0305261}{{\ttfamily arXiv:hep-ph/0305261
  [hep-ph]}}.

\bibitem{Pospelov:2008zw}
M.~Pospelov, ``{Secluded U(1) below the weak scale},''
  \href{http://dx.doi.org/10.1103/PhysRevD.80.095002}{{\em Phys. Rev.}
  {\bfseries D80} (2009) 095002},
\href{http://arxiv.org/abs/0811.1030}{{\ttfamily arXiv:0811.1030 [hep-ph]}}.

\bibitem{Lees:2014xha}
{\bfseries BaBar} Collaboration, J.~P. Lees {\em et al.}, ``{Search for a Dark
  Photon in $e^+e^-$ Collisions at BaBar},''
  \href{http://dx.doi.org/10.1103/PhysRevLett.113.201801}{{\em Phys. Rev.
  Lett.} {\bfseries 113} (2014) 201801},
\href{http://arxiv.org/abs/1406.2980}{{\ttfamily arXiv:1406.2980 [hep-ex]}}.

\bibitem{Lees:2017lec}
{\bfseries BaBar} Collaboration, J.~P. Lees {\em et al.}, ``{Search for
  invisible decays of a dark photon produced in e+e- collisions at BaBar},''
\href{http://arxiv.org/abs/1702.03327}{{\ttfamily arXiv:1702.03327 [hep-ex]}}.

\bibitem{Angloher:2015ewa}
{\bfseries CRESST} Collaboration, G.~Angloher {\em et al.}, ``{Results on light
  dark matter particles with a low-threshold CRESST-II detector},''
  \href{http://dx.doi.org/10.1140/epjc/s10052-016-3877-3}{{\em Eur. Phys. J.}
  {\bfseries C76} (2016) 25},
\href{http://arxiv.org/abs/1509.01515}{{\ttfamily arXiv:1509.01515
  [astro-ph.CO]}}.

\bibitem{Batell:2014mga}
B.~Batell, R.~Essig, and Z.~Surujon, ``{Strong Constraints on Sub-GeV Dark
  Sectors from SLAC Beam Dump E137},''
  \href{http://dx.doi.org/10.1103/PhysRevLett.113.171802}{{\em Phys. Rev.
  Lett.} {\bfseries 113} (2014) 171802},
\href{http://arxiv.org/abs/1406.2698}{{\ttfamily arXiv:1406.2698 [hep-ph]}}.

\bibitem{Izaguirre:2017bqb}
E.~Izaguirre, Y.~Kahn, G.~Krnjaic, and M.~Moschella, ``{Testing Light Dark
  Matter Coannihilation With Fixed-Target Experiments},''
\href{http://arxiv.org/abs/1703.06881}{{\ttfamily arXiv:1703.06881 [hep-ph]}}.

\bibitem{Athanassopoulos:1996ds}
{\bfseries LSND} Collaboration, C.~Athanassopoulos {\em et al.}, ``{The Liquid
  scintillator neutrino detector and LAMPF neutrino source},''
  \href{http://dx.doi.org/10.1016/S0168-9002(96)01155-2}{{\em Nucl. Instrum.
  Meth.} {\bfseries A388} (1997) 149--172},
\href{http://arxiv.org/abs/nucl-ex/9605002}{{\ttfamily arXiv:nucl-ex/9605002
  [nucl-ex]}}.

\bibitem{deNiverville:2011it}
P.~deNiverville, M.~Pospelov, and A.~Ritz, ``{Observing a light dark matter
  beam with neutrino experiments},''
  \href{http://dx.doi.org/10.1103/PhysRevD.84.075020}{{\em Phys. Rev.}
  {\bfseries D84} (2011) 075020},
\href{http://arxiv.org/abs/1107.4580}{{\ttfamily arXiv:1107.4580 [hep-ph]}}.

\bibitem{deNiverville:2016rqh}
P.~deNiverville, C.-Y. Chen, M.~Pospelov, and A.~Ritz, ``{Light dark matter in
  neutrino beams: production modelling and scattering signatures at MiniBooNE,
  T2K and SHiP},'' \href{http://dx.doi.org/10.1103/PhysRevD.95.035006}{{\em
  Phys. Rev.} {\bfseries D95} no.~3, (2017) 035006},
\href{http://arxiv.org/abs/1609.01770}{{\ttfamily arXiv:1609.01770 [hep-ph]}}.

\bibitem{Banerjee:2016tad}
{\bfseries NA64} Collaboration, D.~Banerjee {\em et al.}, ``{Search for
  invisible decays of sub-GeV dark photons in missing-energy events at the CERN
  SPS},'' \href{http://dx.doi.org/10.1103/PhysRevLett.118.011802}{{\em Phys.
  Rev. Lett.} {\bfseries 118} (2017) 011802},
\href{http://arxiv.org/abs/1610.02988}{{\ttfamily arXiv:1610.02988 [hep-ex]}}.

\bibitem{Battaglieri:2016ggd}
{\bfseries BDX} Collaboration, M.~Battaglieri {\em et al.}, ``{Dark Matter
  Search in a Beam-Dump eXperiment (BDX) at Jefferson Lab},''
\href{http://arxiv.org/abs/1607.01390}{{\ttfamily arXiv:1607.01390 [hep-ex]}}.

\bibitem{Bondi:2017gul}
{\bfseries BDX} Collaboration, M.~Bondí, ``{Light Dark Matter search in a
  beam-dump experiment: BDX at Jefferson Lab},''
\href{http://dx.doi.org/10.1051/epjconf/201714201005}{{\em EPJ Web Conf.}
  {\bfseries 142} (2017) 01005}.

\bibitem{HeartyBelle2}
C.~Hearty, ``Dark Sector Searches at B-factories and outlook for Belle II,''
  2017.
\newblock
  \url{https://indico.fnal.gov/getFile.py/access?contribId=123&sessionId=9&resId=0&materialId=slides&confId=13702}.

\bibitem{CoherentWeb}
``\rm {COHERENT} experiment.'' \url{http://sites.duke.edu/coherent}.

\bibitem{deNiverville:2015mwa}
P.~deNiverville, M.~Pospelov, and A.~Ritz, ``{Light new physics in coherent
  neutrino-nucleus scattering experiments},''
  \href{http://dx.doi.org/10.1103/PhysRevD.92.095005}{{\em Phys. Rev.}
  {\bfseries D92} (2015) 095005},
\href{http://arxiv.org/abs/1505.07805}{{\ttfamily arXiv:1505.07805 [hep-ph]}}.

\bibitem{Balewski:2014pxa}
J.~Balewski {\em et al.}, ``{The DarkLight Experiment: A Precision Search for
  New Physics at Low Energies},''
\newblock 2014.
\newblock
\href{http://arxiv.org/abs/1412.4717}{{\ttfamily arXiv:1412.4717
  [physics.ins-det]}}.
\newblock

\bibitem{LDMXWeb}
``\rm {LDMX} experiment.''
  \url{https://confluence.slac.stanford.edu/display/MME/Light+Dark+Matter+Experiment}.

\bibitem{Aguilar-Arevalo:2017mqx}
{\bfseries MiniBooNE} Collaboration, A.~A. Aguilar-Arevalo {\em et al.},
  ``{Dark Matter Search in a Proton Beam Dump with MiniBooNE},''
  \href{http://dx.doi.org/10.1103/PhysRevLett.118.221803}{{\em Phys. Rev.
  Lett.} {\bfseries 118} (2017) 221803},
\href{http://arxiv.org/abs/1702.02688}{{\ttfamily arXiv:1702.02688 [hep-ex]}}.

\bibitem{MMAPS}
J.~Alexander, ``{Dark Photon Search in e+e- Annihilation},''
\newblock 2016.
\newblock
  \url{https://indico.cern.ch/event/507783/contributions/2150181/attachments/1266367/1874844/SLAC-MMAPS-alexander.pdf}.

\bibitem{na64future}
S.~N. Gninenko, ``{NA64$++$},''
\newblock 2017.
\newblock
  \url{https://indico.cern.ch/event/608491/contributions/2457799/attachments/1420197/2176021/NA64_0317.pdf}.

\bibitem{Raggi:2014zpa}
M.~Raggi and V.~Kozhuharov, ``{Proposal to Search for a Dark Photon in Positron
  on Target Collisions at DA$\Phi$NE Linac},''
  \href{http://dx.doi.org/10.1155/2014/959802}{{\em Adv. High Energy Phys.}
  {\bfseries 2014} (2014) 959802},
\href{http://arxiv.org/abs/1403.3041}{{\ttfamily arXiv:1403.3041
  [physics.ins-det]}}.

\bibitem{Raggi:2015gza}
M.~Raggi, V.~Kozhuharov, and P.~Valente, ``{The PADME experiment at LNF},''
  \href{http://dx.doi.org/10.1051/epjconf/20159601025}{{\em EPJ Web Conf.}
  {\bfseries 96} (2015) 01025},
\href{http://arxiv.org/abs/1501.01867}{{\ttfamily arXiv:1501.01867 [hep-ex]}}.

\bibitem{Alekhin:2015byh}
S.~Alekhin {\em et al.}, ``{A facility to Search for Hidden Particles at the
  CERN SPS: the SHiP physics case},''
  \href{http://dx.doi.org/10.1088/0034-4885/79/12/124201}{{\em Rept. Prog.
  Phys.} {\bfseries 79} (2016) 124201},
\href{http://arxiv.org/abs/1504.04855}{{\ttfamily arXiv:1504.04855 [hep-ph]}}.

\bibitem{Anelli:2015pba}
{\bfseries SHiP} Collaboration, M.~Anelli {\em et al.}, ``{A facility to Search
  for Hidden Particles (SHiP) at the CERN SPS},''
\href{http://arxiv.org/abs/1504.04956}{{\ttfamily arXiv:1504.04956
  [physics.ins-det]}}.

\bibitem{Antonello:2015lea}
{\bfseries LAr1-ND, ICARUS-WA104, MicroBooNE} Collaboration, M.~Antonello {\em
  et al.}, ``{A Proposal for a Three Detector Short-Baseline Neutrino
  Oscillation Program in the Fermilab Booster Neutrino Beam},''
\href{http://arxiv.org/abs/1503.01520}{{\ttfamily arXiv:1503.01520
  [physics.ins-det]}}.

\bibitem{SBNRVW}
R.~Van~de Water, ``{Future Sub-GeV Dark Matter Searches with Proton Fixed
  Targets at FNAL}.''
  \url{https://indico.fnal.gov/getFile.py/access?contribId=157&sessionId=6&resId=10&materialId=minutes&confId=13702}.

\bibitem{Wojtsekhowski:2012zq}
B.~Wojtsekhowski, D.~Nikolenko, and I.~Rachek, ``{Searching for a new force at
  VEPP-3},''
\href{http://arxiv.org/abs/1207.5089}{{\ttfamily arXiv:1207.5089 [hep-ex]}}.

\bibitem{Izaguirre:2015yja}
E.~Izaguirre, G.~Krnjaic, P.~Schuster, and N.~Toro, ``{Analyzing the Discovery
  Potential for Light Dark Matter},''
  \href{http://dx.doi.org/10.1103/PhysRevLett.115.251301}{{\em Phys. Rev.
  Lett.} {\bfseries 115} (2015) 251301},
\href{http://arxiv.org/abs/1505.00011}{{\ttfamily arXiv:1505.00011 [hep-ph]}}.

\bibitem{Buschmann:2015awa}
M.~Buschmann, J.~Kopp, J.~Liu, and P.~A.~N. Machado, ``{Lepton Jets from
  Radiating Dark Matter},''
  \href{http://dx.doi.org/10.1007/JHEP07(2015)045}{{\em JHEP} {\bfseries 07}
  (2015) 045},
\href{http://arxiv.org/abs/1505.07459}{{\ttfamily arXiv:1505.07459 [hep-ph]}}.

\bibitem{TuckerSmith:2001hy}
D.~Tucker-Smith and N.~Weiner, ``{Inelastic dark matter},''
  \href{http://dx.doi.org/10.1103/PhysRevD.64.043502}{{\em Phys. Rev.}
  {\bfseries D64} (2001) 043502},
\href{http://arxiv.org/abs/hep-ph/0101138}{{\ttfamily arXiv:hep-ph/0101138
  [hep-ph]}}.

\bibitem{Izaguirre:2015zva}
E.~Izaguirre, G.~Krnjaic, and B.~Shuve, ``{Discovering Inelastic Thermal-Relic
  Dark Matter at Colliders},''
  \href{http://dx.doi.org/10.1103/PhysRevD.93.063523}{{\em Phys. Rev.}
  {\bfseries D93} (2016) 063523},
\href{http://arxiv.org/abs/1508.03050}{{\ttfamily arXiv:1508.03050 [hep-ph]}}.

\bibitem{Gondolo:1990dk}
P.~Gondolo and G.~Gelmini, ``{Cosmic abundances of stable particles: Improved
  analysis},''
\href{http://dx.doi.org/10.1016/0550-3213(91)90438-4}{{\em Nucl. Phys.}
  {\bfseries B360} (1991) 145--179}.

\bibitem{Griest:1990kh}
K.~Griest and D.~Seckel, ``{Three exceptions in the calculation of relic
  abundances},''
\href{http://dx.doi.org/10.1103/PhysRevD.43.3191}{{\em Phys. Rev.} {\bfseries
  D43} (1991) 3191--3203}.

\bibitem{scipy}
E.~Jones, T.~Oliphant, P.~Peterson, {\em et al.}, ``{SciPy}: Open source
  scientific tools for {Python},'' 2001--.
\newblock \url{http://www.scipy.org}. [Online; accessed 6/10/2017].

\bibitem{Zaghloul:2012:A9C:2049673.2049679}
M.~R. Zaghloul and A.~N. Ali, ``Algorithm 916: Computing the Faddeyeva and
  Voigt Functions,'' \href{http://dx.doi.org/10.1145/2049673.2049679}{{\em ACM
  Trans. Math. Softw.} {\bfseries 38} (Jan., 2012) 15:1--15:22}.

\bibitem{Battaglieri:2017aum}
M.~Battaglieri {\em et al.}, ``{US Cosmic Visions: New Ideas in Dark Matter
  2017: Community Report},''
\href{http://arxiv.org/abs/1707.04591}{{\ttfamily arXiv:1707.04591 [hep-ph]}}.

\bibitem{Essig:2011nj}
R.~Essig, J.~Mardon, and T.~Volansky, ``{Direct Detection of Sub-GeV Dark
  Matter},'' \href{http://dx.doi.org/10.1103/PhysRevD.85.076007}{{\em Phys.
  Rev.} {\bfseries D85} (2012) 076007},
\href{http://arxiv.org/abs/1108.5383}{{\ttfamily arXiv:1108.5383 [hep-ph]}}.

\bibitem{Essig:2015cda}
R.~Essig, M.~Fernandez-Serra, J.~Mardon, A.~Soto, T.~Volansky, and T.-T. Yu,
  ``{Direct Detection of sub-GeV Dark Matter with Semiconductor Targets},''
  \href{http://dx.doi.org/10.1007/JHEP05(2016)046}{{\em JHEP} {\bfseries 05}
  (2016) 046},
\href{http://arxiv.org/abs/1509.01598}{{\ttfamily arXiv:1509.01598 [hep-ph]}}.

\bibitem{Essig:2012yx}
R.~Essig, A.~Manalaysay, J.~Mardon, P.~Sorensen, and T.~Volansky, ``{First
  Direct Detection Limits on sub-GeV Dark Matter from XENON10},''
  \href{http://dx.doi.org/10.1103/PhysRevLett.109.021301}{{\em Phys. Rev.
  Lett.} {\bfseries 109} (2012) 021301},
\href{http://arxiv.org/abs/1206.2644}{{\ttfamily arXiv:1206.2644
  [astro-ph.CO]}}.

\bibitem{Essig:2017kqs}
R.~Essig, T.~Volansky, and T.-T. Yu, ``{New Constraints and Prospects for
  sub-GeV Dark Matter Scattering off Electrons in Xenon},''
\href{http://arxiv.org/abs/1703.00910}{{\ttfamily arXiv:1703.00910 [hep-ph]}}.

\bibitem{Derenzo:2016fse}
S.~Derenzo, R.~Essig, A.~Massari, A.~Soto, and T.-T. Yu, ``{Direct Detection of
  sub-GeV Dark Matter with Scintillating Targets},''
  \href{http://dx.doi.org/10.1103/PhysRevD.96.016026}{{\em Phys. Rev.}
  {\bfseries D96} (2017) 016026},
\href{http://arxiv.org/abs/1607.01009}{{\ttfamily arXiv:1607.01009 [hep-ph]}}.

\bibitem{Hochberg:2016ntt}
Y.~Hochberg, Y.~Kahn, M.~Lisanti, C.~G. Tully, and K.~M. Zurek, ``{Directional
  detection of dark matter with two-dimensional targets},''
  \href{http://dx.doi.org/10.1016/j.physletb.2017.06.051}{{\em Phys. Lett.}
  {\bfseries B772} (2017) 239--246},
\href{http://arxiv.org/abs/1606.08849}{{\ttfamily arXiv:1606.08849 [hep-ph]}}.

\bibitem{Hochberg:2015fth}
Y.~Hochberg, M.~Pyle, Y.~Zhao, and K.~M. Zurek, ``{Detecting Superlight Dark
  Matter with Fermi-Degenerate Materials},''
  \href{http://dx.doi.org/10.1007/JHEP08(2016)057}{{\em JHEP} {\bfseries 08}
  (2016) 057},
\href{http://arxiv.org/abs/1512.04533}{{\ttfamily arXiv:1512.04533 [hep-ph]}}.

\bibitem{Hochberg:2015pha}
Y.~Hochberg, Y.~Zhao, and K.~M. Zurek, ``{Superconducting Detectors for
  Superlight Dark Matter},''
  \href{http://dx.doi.org/10.1103/PhysRevLett.116.011301}{{\em Phys. Rev.
  Lett.} {\bfseries 116} (2016) 011301},
\href{http://arxiv.org/abs/1504.07237}{{\ttfamily arXiv:1504.07237 [hep-ph]}}.

\bibitem{Aguilar-Arevalo:2016ndq}
{\bfseries DAMIC} Collaboration, A.~Aguilar-Arevalo {\em et al.}, ``{Search for
  low-mass WIMPs in a 0.6 kg day exposure of the DAMIC experiment at SNOLAB},''
  \href{http://dx.doi.org/10.1103/PhysRevD.94.082006}{{\em Phys. Rev.}
  {\bfseries D94} (2016) 082006},
\href{http://arxiv.org/abs/1607.07410}{{\ttfamily arXiv:1607.07410
  [astro-ph.CO]}}.

\bibitem{Aguilar-Arevalo:2016zop}
{\bfseries DAMIC} Collaboration, A.~Aguilar-Arevalo {\em et al.}, ``{First
  Direct-Detection Constraints on eV-Scale Hidden-Photon Dark Matter with DAMIC
  at SNOLAB},'' \href{http://dx.doi.org/10.1103/PhysRevLett.118.141803}{{\em
  Phys. Rev. Lett.} {\bfseries 118} (2017) 141803},
\href{http://arxiv.org/abs/1611.03066}{{\ttfamily arXiv:1611.03066
  [astro-ph.CO]}}.

\bibitem{Tiffenberg:2017aac}
J.~Tiffenberg, M.~Sofo-Haro, A.~Drlica-Wagner, R.~Essig, Y.~Guardincerri,
  S.~Holland, T.~Volansky, and T.-T. Yu, ``{Single-electron and single-photon
  sensitivity with a silicon Skipper CCD},''
\href{http://arxiv.org/abs/1706.00028}{{\ttfamily arXiv:1706.00028
  [physics.ins-det]}}.

\bibitem{Agnese:2014aze}
{\bfseries SuperCDMS} Collaboration, R.~Agnese {\em et al.}, ``{Search for
  Low-Mass Weakly Interacting Massive Particles with SuperCDMS},''
  \href{http://dx.doi.org/10.1103/PhysRevLett.112.241302}{{\em Phys. Rev.
  Lett.} {\bfseries 112} (2014) 241302},
\href{http://arxiv.org/abs/1402.7137}{{\ttfamily arXiv:1402.7137 [hep-ex]}}.

\bibitem{Agnese:2015nto}
{\bfseries SuperCDMS} Collaboration, R.~Agnese {\em et al.}, ``{New Results
  from the Search for Low-Mass Weakly Interacting Massive Particles with the
  CDMS Low Ionization Threshold Experiment},''
  \href{http://dx.doi.org/10.1103/PhysRevLett.116.071301}{{\em Phys. Rev.
  Lett.} {\bfseries 116} (2016) 071301},
\href{http://arxiv.org/abs/1509.02448}{{\ttfamily arXiv:1509.02448
  [astro-ph.CO]}}.

\bibitem{Schutz:2016tid}
K.~Schutz and K.~M. Zurek, ``{Detectability of Light Dark Matter with
  Superfluid Helium},''
  \href{http://dx.doi.org/10.1103/PhysRevLett.117.121302}{{\em Phys. Rev.
  Lett.} {\bfseries 117} (2016) 121302},
\href{http://arxiv.org/abs/1604.08206}{{\ttfamily arXiv:1604.08206 [hep-ph]}}.

\bibitem{Knapen:2016cue}
S.~Knapen, T.~Lin, and K.~M. Zurek, ``{Light Dark Matter in Superfluid Helium:
  Detection with Multi-excitation Production},''
  \href{http://dx.doi.org/10.1103/PhysRevD.95.056019}{{\em Phys. Rev.}
  {\bfseries D95} (2017) 056019},
\href{http://arxiv.org/abs/1611.06228}{{\ttfamily arXiv:1611.06228 [hep-ph]}}.

\bibitem{Guo:2013dt}
W.~Guo and D.~N. McKinsey, ``{Concept for a dark matter detector using liquid
  helium-4},'' \href{http://dx.doi.org/10.1103/PhysRevD.87.115001}{{\em Phys.
  Rev.} {\bfseries D87} (2013) 115001},
\href{http://arxiv.org/abs/1302.0534}{{\ttfamily arXiv:1302.0534
  [astro-ph.IM]}}.

\bibitem{Car:2017}
F.~W. Carter, S.~A. Hertel, M.~J. Rooks, D.~N. McKinsey, and D.~E. Prober,
  ``Toward the In Situ Detection of Individual He2 Excimers Using a Ti TES in
  Superfluid Helium,'' {\em IEEE Transactions an Applied Superconductivity}
  {\bfseries 25} (2015) 2100707.

\bibitem{Maris:2017xvi}
H.~J. Maris, G.~M. Seidel, and D.~Stein, ``{Dark Matter Detection Using Helium
  Evaporation and Field Ionization},''
\href{http://arxiv.org/abs/1706.00117}{{\ttfamily arXiv:1706.00117
  [astro-ph.IM]}}.

\bibitem{Essig:2016crl}
R.~Essig, J.~Mardon, O.~Slone, and T.~Volansky, ``{Detection of sub-GeV Dark
  Matter and Solar Neutrinos via Chemical-Bond Breaking},''
  \href{http://dx.doi.org/10.1103/PhysRevD.95.056011}{{\em Phys. Rev.}
  {\bfseries D95} (2017) 056011},
\href{http://arxiv.org/abs/1608.02940}{{\ttfamily arXiv:1608.02940 [hep-ph]}}.

\bibitem{Budnik:2017sbu}
R.~Budnik, O.~Chesnovsky, O.~Slone, and T.~Volansky, ``{Direct Detection of
  Light Dark Matter and Solar Neutrinos via Color Center Production in
  Crystals},''
\href{http://arxiv.org/abs/1705.03016}{{\ttfamily arXiv:1705.03016 [hep-ph]}}.

\bibitem{Bunting:2017net}
P.~C. Bunting, G.~Gratta, T.~Melia, and S.~Rajendran, ``{Magnetic Bubble
  Chambers and Sub-GeV Dark Matter Direct Detection},''
  \href{http://dx.doi.org/10.1103/PhysRevD.95.095001}{{\em Phys. Rev.}
  {\bfseries D95} (2017) 095001},
\href{http://arxiv.org/abs/1701.06566}{{\ttfamily arXiv:1701.06566 [hep-ph]}}.

\bibitem{Baxter:2017ozv}
D.~Baxter {\em et al.}, ``{First Demonstration of a Scintillating Xenon Bubble
  Chamber for Detecting Dark Matter and Coherent Elastic Neutrino-Nucleus
  Scattering},'' \href{http://dx.doi.org/10.1103/PhysRevLett.118.231301}{{\em
  Phys. Rev. Lett.} {\bfseries 118} (2017) 231301},
\href{http://arxiv.org/abs/1702.08861}{{\ttfamily arXiv:1702.08861
  [physics.ins-det]}}.

\bibitem{Kouvaris:2016afs}
C.~Kouvaris and J.~Pradler, ``{Probing sub-GeV Dark Matter with conventional
  detectors},'' \href{http://dx.doi.org/10.1103/PhysRevLett.118.031803}{{\em
  Phys. Rev. Lett.} {\bfseries 118} (2017) 031803},
\href{http://arxiv.org/abs/1607.01789}{{\ttfamily arXiv:1607.01789 [hep-ph]}}.

\bibitem{McCabe:2017rln}
C.~McCabe, ``{New constraints and discovery potential of sub-GeV dark matter
  with xenon detectors},''
  \href{http://dx.doi.org/10.1103/PhysRevD.96.043010}{{\em Phys. Rev.}
  {\bfseries D96} (2017) 043010},
\href{http://arxiv.org/abs/1702.04730}{{\ttfamily arXiv:1702.04730 [hep-ph]}}.

\bibitem{Servant:2002aq}
G.~Servant and T.~M.~P. Tait, ``{Is the lightest Kaluza-Klein particle a viable
  dark matter candidate?},''
  \href{http://dx.doi.org/10.1016/S0550-3213(02)01012-X}{{\em Nucl. Phys.}
  {\bfseries B650} (2003) 391--419},
\href{http://arxiv.org/abs/hep-ph/0206071}{{\ttfamily arXiv:hep-ph/0206071
  [hep-ph]}}.

\bibitem{Cheng:2002ej}
H.-C. Cheng, J.~L. Feng, and K.~T. Matchev, ``{Kaluza-Klein dark matter},''
  \href{http://dx.doi.org/10.1103/PhysRevLett.89.211301}{{\em Phys. Rev. Lett.}
  {\bfseries 89} (2002) 211301},
\href{http://arxiv.org/abs/hep-ph/0207125}{{\ttfamily arXiv:hep-ph/0207125
  [hep-ph]}}.

\bibitem{Kakizaki:2005en}
M.~Kakizaki, S.~Matsumoto, Y.~Sato, and M.~Senami, ``{Significant effects of
  second KK particles on LKP dark matter physics},''
  \href{http://dx.doi.org/10.1103/PhysRevD.71.123522}{{\em Phys. Rev.}
  {\bfseries D71} (2005) 123522},
\href{http://arxiv.org/abs/hep-ph/0502059}{{\ttfamily arXiv:hep-ph/0502059
  [hep-ph]}}.

\bibitem{Feldman:2008xs}
D.~Feldman, Z.~Liu, and P.~Nath, ``{PAMELA Positron Excess as a Signal from the
  Hidden Sector},'' \href{http://dx.doi.org/10.1103/PhysRevD.79.063509}{{\em
  Phys. Rev.} {\bfseries D79} (2009) 063509},
\href{http://arxiv.org/abs/0810.5762}{{\ttfamily arXiv:0810.5762 [hep-ph]}}.

\bibitem{Ibe:2008ye}
M.~Ibe, H.~Murayama, and T.~T. Yanagida, ``{Breit-Wigner Enhancement of Dark
  Matter Annihilation},''
  \href{http://dx.doi.org/10.1103/PhysRevD.79.095009}{{\em Phys. Rev.}
  {\bfseries D79} (2009) 095009},
\href{http://arxiv.org/abs/0812.0072}{{\ttfamily arXiv:0812.0072 [hep-ph]}}.

\bibitem{Guo:2009aj}
W.-L. Guo and Y.-L. Wu, ``{Enhancement of Dark Matter Annihilation via
  Breit-Wigner Resonance},''
  \href{http://dx.doi.org/10.1103/PhysRevD.79.055012}{{\em Phys. Rev.}
  {\bfseries D79} (2009) 055012},
\href{http://arxiv.org/abs/0901.1450}{{\ttfamily arXiv:0901.1450 [hep-ph]}}.

\end{thebibliography}\endgroup

\end{document}